\documentclass[prl,aps,superscriptaddress,notitlepage]{revtex4}
\usepackage{bm}
\usepackage[dvipsnames]{xcolor}
\usepackage[breaklinks,bookmarks = false,pdfpagemode = UseNone, colorlinks= true,]{hyperref}
\hypersetup{linkcolor=RubineRed,citecolor=RoyalBlue,filecolor=Mulberry,urlcolor=RoyalBlue}
\usepackage{times}
\usepackage{latexsym}
\usepackage{graphicx}
\usepackage{microtype}

\usepackage{dcolumn}
\usepackage{soul}
\usepackage{color}
\usepackage{enumerate} 
\newcommand{\stoica}[1]{\textcolor{blue}{#1}}

\begin{document}

\title{Polar-vortex-driven interfacial strain coupling in PbTiO$_3$/SrRuO$_3$ Heterostructures}
\author{S. A. Raza}
\email{asyed@anl.gov}
\email{adnanphyqau@gmail.com}
\affiliation{%
X-ray Science Division, Argonne National Laboratory, Argonne, Illinois 60439, USA}
\author{V. A. Stoica}
\affiliation{%
Department of Materials Science and Engineering, Pennsylvania State University, University Park, Pennsylvania 16802, USA}
\author{H. Zheng}
\affiliation{%
X-ray Science Division, Argonne National Laboratory, Argonne, Illinois 60439, USA}
\author{H. G. Lee}
\affiliation{%
Department of Materials Science and Engineering, University of California, Berkeley, California 94720, United States}
\author{A. Ross}
\affiliation{%
Department of Materials Science and Engineering, Pennsylvania State University, University Park, Pennsylvania 16802, USA}
\author{U. Saha}
\affiliation{%
Department of Materials Science and Engineering, Pennsylvania State University, University Park, Pennsylvania 16802, USA}
\author{L. Q. Chen}
\affiliation{%
Department of Materials Science and Engineering, Pennsylvania State University, University Park, Pennsylvania 16802, USA}
\author{L. W. Martin}
\affiliation{%
 Departments of Materials Science and NanoEngineering, Chemistry, and Physics and Astronomy and the Rice Advanced Materials Institute, Rice University, Houston, Texas 77005, United States}
\author{V. Gopalan}
\affiliation{%
Department of Materials Science and Engineering, Pennsylvania State University, University Park, Pennsylvania 16802, USA}
\author{J. W.\ Freeland}
\email{freeland@anl.gov}
\affiliation{%
X-ray Science Division, Argonne National Laboratory, Argonne, Illinois 60439, USA}

\date{\today}

\begin{abstract}
Interfacial coupling in oxide heterostructures is a central problem in condensed-matter physics, as it typically emerges at the atomic scale through local interactions mediated by lattice polarization and strain. In this work, we investigate nanoscale polar-supertexture-driven interfacial strain coupling in (PbTiO$_3$)$_{16}$/(SrRuO$_3$)$_9$/(PbTiO$_3$)$_{16}$ heterostructures grown on DyScO$_3$(110) substrates. Under appropriate epitaxial strain conditions, the PbTiO$_3$ layers form polar vortex superstructures with a periodicity of approximately 10 nm. We demonstrate that the resulting in-plane nanoscale strain modulation propagates into the SrRuO$_3$ layer. Using element-specific resonant X-ray reflectivity, we probe the nanoscale strain modulations of the strontium and ruthenium sublattices at the interface, revealing strong interfacial strain coupling between the ferroelectric and ferromagnetic layers. These findings provide new insights into engineering nanoscale magnetic modulations through interfacial strain and polarization control.

\stoica{}
%\begin{abstract} \end{abstract}

\end{abstract}
\maketitle
\section{I. Introduction}

The interfaces between complex oxides host emergent phenomena arising from strong coupling among lattice, charge, spin, and orbital degrees of freedom, leading to a broad spectrum of functional properties \cite{zubko2011interface,10.1093/acprof:oso/9780198507789.001.0001,das2020new,hao2025interface}. In particular, oxide heterostructures provide a versatile platform for engineering novel states of matter through interfacial coupling, symmetry breaking, and strain control. When combining different uniform quantum-material layers in heterostructures, these systems enable precise manipulation of electronic, magnetic, and topological properties at the atomic scale \cite{gu2020interfacial,huang2018interface, https://doi.org/10.1002/adma.202106909}.
Moreover, these systems allow precise tuning and control of interfacial material properties in the presence of structural and electronic heterogeneity \cite{junquera2023topological}. 
Among these, magnetoelectric heterostructures formed at ferroelectric (FE)/ferromagnetic (FM) oxide interfaces have attracted significant attention, as they allow for direct coupling between electric polarization and magnetization, offering both a fertile ground for exploring coupled order parameters and a pathway toward low-power, multifunctional device concepts \cite{https://doi.org/10.1002/adma.201502824,taniyama2024artificial,vaz2021epitaxial}. 

Polar topological textures in oxide heterostructures have been extensively explored in (PbTiO$_3$)$_n$/(SrTiO$_3$)$_m$ superlattice systems, where polar ordering stabilizes nanoscale polarization patterns \cite{tang2021periodic, stoica2019optical}. 
Depending on the number of layer repetitions ($n$, \textit{m}; the number of unit cells in each layer), and the substrate on which they are grown, atomic-scale polar configurations such as $a_1/a_2$ ferroelectric domains (\textit{i.e.}, two types of \textit{a} domains with the polar axis parallel to the growth plane), vortices, skyrmions, and mixed phases have so far been realized primarily in spatially confined PbTiO$_3$ layers embedded in carefully designed (PbTiO$_3$)$_n$/(SrTiO$_3$)$_m$ superlattices \cite{https://doi.org/10.1002/adma.201901014, xue2025observation, das2018perspective, das2019observation}.
Extending this platform by replacing the nonmagnetic spacer with a functional magnetic oxide such as SrRuO$_3$ opens a new and promising coupling pathway through lattice-mediated strain transfer~\cite{lichtensteiger2023mapping}. As a correlated $4d$-electron ferromagnet, SrRuO$_3$ exhibits an orthorhombic \textit{Pnma} crystal symmetry and electronic structure and magnetic anisotropy that are highly responsive to epitaxial strain and oxygen-octahedral distortions.
This exceptional sensitivity makes SrRuO$_3$ an ideal medium through which polar-driven lattice modulations in adjacent PbTiO$_3$ layers can be harnessed to actively tune and control its magnetic ground state in connection with structural changes at interfaces~\cite{zhou2014electric, egoavil2013atomic}. Therefore, understanding the structural modulations at heterostructure interfaces is crucial for building emergent magnetic behavior.

Exploring interfacial coupling between PbTiO$_3$ and SrRuO$_3$ in such heterostructures provides a compelling platform to investigate strain-driven interfacial magnetoelectric coupling. Nanoscale polarization textures in PbTiO$_3$ generate spatially modulated strain fields that can propagate into SrRuO$_3$, inducing local structural distortions and modifying its electronic and magnetic properties. As SrRuO$_3$ simultaneously acts as a metallic (or semiconducting in the ultra-thin limit) ferromagnet and epitaxial electrode, (PbTiO$_3$)$_{16}$/(SrRuO$_3$)$_9$/(PbTiO$_3$)$_{16}$ heterostructures enable efficient strain transmission across atomically sharp interfaces, making them an ideal model system for exploring lattice-controlled ferromagnetism and voltage-tunable spintronic functionality \cite{seddon2021real, velev2016predictive, yao2022ferroelectric}.
Recent theoretical work~\cite{hu2017understanding} suggests two principal routes for coupling polarization and magnetization in such systems: (i) strain-mediated coupling and (ii) polarization-driven modification of the electronic and magnetic structure. The static modulations are expected to generate complex magnetic textures that can enhance the topological-Hall effect, for example. SrRuO$_3$ shows chiral magnetic modulations, although these occur at very large length scales as compared to periodic structures found in (PbTiO$_3$)$_n$/(SrTiO$_3$)$_m$ or PbTiO$_3$ films~\cite{seddon2021real, huang2020detection, abid2021creating}.

Such smaller length scale lattice modulations in contact with SrRuO$_3$ were recently explored by Céline, \textit{et al}. in SrRuO$_3$/PbTiO$_3$/SrRuO$_3$ heterostructures, showing that above a critical ferroelectric thickness, ferroelastic distortions propagate into the adjacent SrRuO$_3$ layers and induce a modulated structural response beyond the ferroelectric boundaries \cite{lichtensteiger2023nanoscale}. Moreover, PbTiO$_3$/SrRuO$_3$ heterostructures can be used to generate ferroelectric supercrystals, chiral magnetic textures, and magnetoelastic effects, enriching the prospects to correlate the structural coupling between dissimilar materials linked to a multitude of ferroic properties \cite{hadjimichael2021metal, seddon2021real, rusu2022ferroelectric}. A key open question, however, remains: is there direct evidence for strain penetration into the SrRuO$_3$ layer?

Here, we investigate polar-vortex-driven interfacial-strain coupling using resonant soft X-ray scattering (RSXS), a noninvasive, element-specific diffraction technique that combines structural sensitivity with spectroscopic selectivity. By tuning to specific absorption edges, RSXS enables sublattice-resolved access to the strain modulated lattice structure, allowing us to disentangle the distinct responses of the strontium and ruthenium sublattices in SrRuO$_3$. This capability is particularly important for SrRuO$_3$, where epitaxial strain modifies the Ru-O bond angles, octahedral rotations, and orbital hybridization, thereby strongly influencing its electronic and magnetic ground state \cite{doi:10.1073/pnas.2101946118, PhysRevLett.102.177601, AdvancedMaterials2021}.  We examine (PbTiO$_3$)$_{16}$/(SrRuO$_3$)$_9$/(PbTiO$_3$)$_{16}$ heterostructures that are epitaxially grown on DyScO$_3$ (110) substrates. Our RSXS measurements demonstrate that the nanoscale, vortex-induced strain modulations in PbTiO$_3$ propagate into the SrRuO$_3$ layer and couple differently to the constituent atomic sublattices, highlighting how nanoscale strain modulations can be propagated across interfaces to create nanoscale electronic and magnetic structure modulations.

\section{II. Results and Discussion}
\label{Results and Discussion}

\subsection*{A. Sample growth}
Trilayers of (PbTiO$_3$)$_{16}$/(SrRuO$_3$)$_9$/(PbTiO$_3$)$_{16}$ (where 16 and 9 denote the thicknesses of the PbTiO$_3$ and SrRuO$_3$ layers in terms of unit cells) were fabricated on 2 unit cells (\,u.c.)\ SrTiO$_3$-buffered DyScO$_3$ (110) substrates using pulsed-laser deposition using a KrF excimer laser (Coherent, LPX-300; $\lambda = 248\,\mathrm{nm}$). The SrTiO$_3$-buffer layer was first grown at a substrate temperature of $690\,^{\circ}\mathrm{C}$ and an oxygen pressure of 80\,mTorr. Next, the substrate temperature was reduced to $600\,^{\circ}\mathrm{C}$ and the oxygen pressure increased to 200\,mTorr for the subsequent alternating growth of the PbTiO$_3$ and SrRuO$_3$ layers. The laser energy was 400 and $360\,\mathrm{mJ\,pulse^{-1}}$ for the growth of the SrTiO$_3$ and trilayer structure, respectively. Optimizing the laser energy was essential for stabilizing polar vortices within the heterostructures. The thicknesses of the SrTiO$_3$, PbTiO$_3$, and SrRuO$_3$ layers were precisely controlled at 2, 16, and 9 unit cells, respectively, by adjusting the deposition time. Following growth, the (PbTiO$_3$)$_{16}$/(SrRuO$_3$)$_9$/(PbTiO$_3$)$_{16}$ films were slowly cooled to room temperature under an oxygen pressure of 200\,mTorr.

\subsection*{B. Structural and electronic properties}
\begin{figure}[h]
\centering
\includegraphics[width=.75\textwidth]{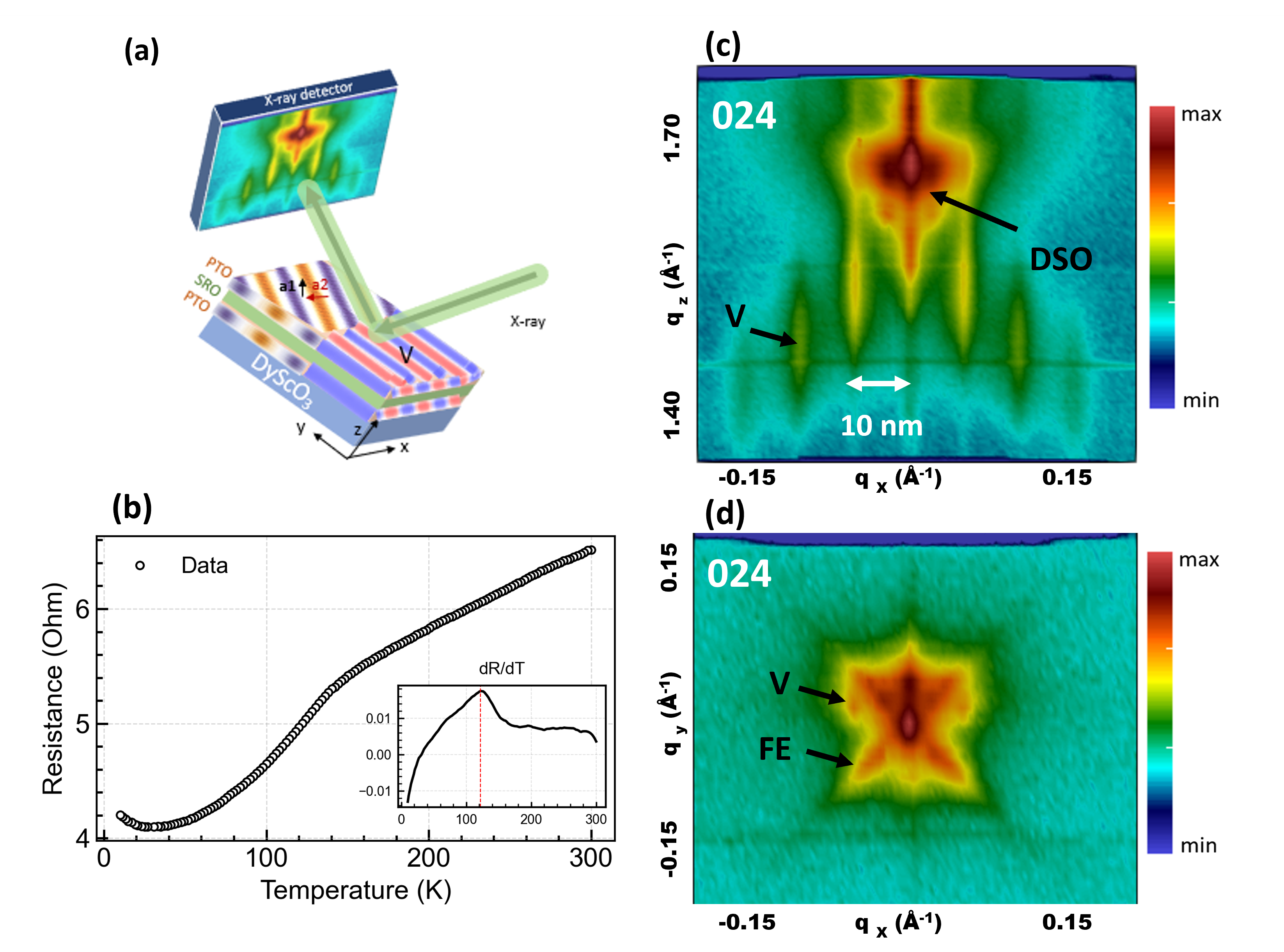} 
\caption{\label{Fig1} Schematic of the X-ray diffraction geometry for (PbTiO$_3$)$_{16}$/(SrRuO$_3$)$_9$/(PbTiO$_3$)$_{16}$ trilayers (a). Electrical resistance as a function of temperature measured upon cooling the sample from 300 to 10~K with the current applied along the [001] (b); the inset shows the corresponding $dR/dT$. The dashed line marks the ferromagnetic Curie temperature ($T_\mathrm{C}$) of the sample at $T = 125$~K, derived from the peak shown in the inset. Hard-X-ray reciprocal space mapping the xz planar cut intersecting the 024 Bragg peak $q_{[00L]}$ (c), and $q_{[H00]}$ (d), showing the marked reflections of the two-phase mixture of in-plane ferroelectric–ferroelastic $a_1/a_2$ domains (FE) and polar vortex (V).
}
\end{figure}
To investigate the interfacial structural coupling in PbTiO$_3$/SrRuO$_3$ heterostructures, we employ non-resonant hard X-ray diffraction measurements at beamline 6-ID-B of the Advanced Photon Source (APS), Argonne National Laboratory in combination with transport measurements at variable temperature (Fig.~\ref{Fig1}). These measurements enabled detailed static structural characterization of the lattice parameters and structural quality of the heterostructures in addition to electronic transport information. All measurements were carried out at room temperature using an incident X-ray energy of 11.215~keV. Three-dimensional reciprocal space maps (RSMs) of the scattered X-ray intensity from the thin films were collected using a LAMBDA area detector with a pixel size of 55~$\mu$m. 
Figure~\ref{Fig1}(a) schematically illustrates the hard X-ray diffraction geometry for the (PbTiO$_3$)$_{16}$/(SrRuO$_3$)$_9$/(PbTiO$_3$)$_{16}$ trilayer grown on DyScO$_3$ (110) substrates, along with the corresponding diffraction recorded by an area detector. The observed scattering arises from the coexistence of $a_1/a_2$ ferroelastic domains and polar-vortex structures, highlighting a complex nanoscale phase mixture within the heterostructure at room temperature.

Figure~\ref{Fig1}(b) shows the temperature dependence of the electrical resistance. Over the full temperature range studied, the transport behavior is characteristic of a SrRuO$_3$ layer that has a much larger conductivity as compared to the PbTiO$_3$ layers and the insulating DyScO$_3$ substrate. The inset displays the temperature derivative of the resistance. The dashed line marks $T = 125$~K, corresponding to the ferromagnetic Curie temperature ($T_\mathrm{C}$) of this sample. Below approximately 125~K, the resistance in the magnetic state follows a functional form consistent with Fermi-liquid behavior, while above 125~K, the resistance increases nearly linearly with temperature, indicative of bad-metal transport in the paramagnetic phases. This allows us to conclude that the SrRuO$_3$ spacer orders ferromagnetically at a slightly reduced T$_c$ as compared to bulk~\cite{DWIVEDI2025141963, dwivedi2019role, RevModPhys.84.253}.

Hard X-ray reciprocal space maps (3D-RSMs) were reconstructed from experimental rocking curve scans about the 024$_\mathrm{pc}$-diffraction condition of the DyScO$_3$ substrate; Fig.~\ref{Fig1}(c) and (d) display representative planar slices extracted along the [00L]$_\mathrm{pc}$ and [-H,H,0]$_\mathrm{pc}$, respectively. These data reveal the film's Bragg peaks and thickness fringes along the $q_z$ direction. The positions of the vortex-satellite reflections quantitatively confirm a mixed-phase state consisting of alternating $a_1/a_2$ ferroelectric domains and polar vortices in the PbTiO$_3$ layers, with an out-of-plane lattice constant of 3.907~\AA~and~3.93~\AA, respectively, consistent with previous reports on phase mixed structures in (PbTiO$_3$)$_n$/(SrTiO$_3$)$_m$ superlattices~\cite{cha2024quasi, hong2017stability, damodaran2017phase}. In the vortex region, the ferroelectric polarization in the PbTiO$_3$ layers forms pairs of clockwise and counterclockwise polar vortices arranged in a periodic lattice, as reported previously \cite{hong2017stability, yadav2016observation}, arising from the competition between electrostatic energy and interfacial coupling.
The vortices extend as tube-like features along the [010]$_\mathrm{pc}$ and arrange into a periodic array along the [001]$_\mathrm{pc}$ (Fig.~\ref{Fig1}(a)). The out-of-plane superlattice thickness is determined to be 16.5~nm, corresponding to the total thickness of the heterostructure. Furthermore, as shown (Fig.~\ref{Fig1}(c)), a series of pronounced satellite peaks is observed along the in-plane $\langle100\rangle_\mathrm{pc}$, indicating a vortex periodicity of approximately 10~nm within the PbTiO$_3$ layers. Moreover, the satellite peak at $q[110] \approx  0.045~$\AA$^{-1}$ (marked in the inset) corresponds to an in-plane $a_1/a_2$ ferroelastic domain periodicity of approximately 14~nm (Fig.~\ref{Fig1}(d)). The peak full-width-at-half-maximum, $\Delta q \approx 0.02~$\AA$^{-1}$, reflects a lateral coherence length of $\approx$ 30 nm, highlighting the limited spatial extent of the domain pattern in ultrathin PbTiO$_3$ films on DyScO$_3$ \cite{stoica2019optical, damodaran2017phase}.

\subsection*{C. Resonant Soft X-ray Scattering}
\begin{figure}[h]
\centering
\includegraphics[width=.85\textwidth, height=0.6\textheight, keepaspectratio]{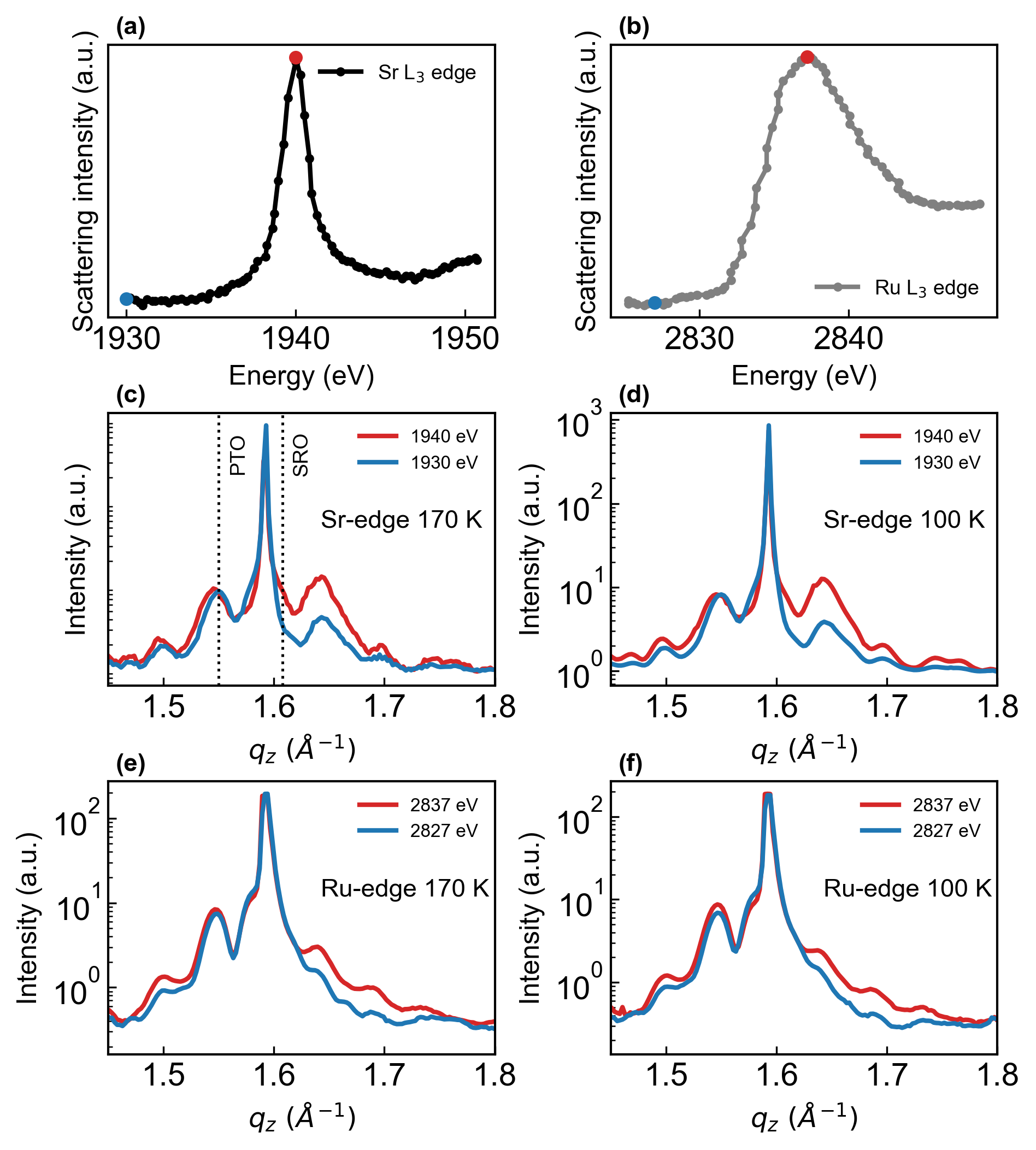} 
\caption{\label{Fig2} Scattering intensity measured at a fixed $q$ (0,0, 1.61) value across the strontium $L_{3}$-edge (a) and the ruthenium $L_{3}$-edge (b); red and blue dots indicate the resonant and off-resonant photon energies. Specular soft X-ray reflectivity at the strontium $L_{3}$-edge measured at resonant and off-resonant energies at 170~K (c) and 100~K (d). Specular $q_z$ scan at the ruthenium $L_{3}$-edge acquired at resonant and off-resonant energies at 170~K (e). Specular $q_z$ scan at the ruthenium $L_{3}$-edge acquired at resonant and off-resonant energies at 100~K (f).
  }
\end{figure}
To understand element-specific contributions to the scattering factor, resonant soft X-ray scattering (RSXS) experiments were performed at the Advanced Photon Source using the scattering endstation at 29-ID-D beamline. The photon energy of the X-ray beam was tuned using an elliptically polarizing undulator, and the polarization was set to right-circularly polarized (RCP) X-rays. The intensity of the X-rays scattered by the sample was measured with an in-vacuum, aluminum-coated silicon photodiode that is sensitive to soft X-rays. The measured scattered intensity was normalized to $I_0$ at each photon energy to account for the energy dependence of the incident photon flux. Resonance profiles were obtained by tuning the photon energy through the strontium-$L_3$ edge (approximately 1930-1950~eV) and the ruthenium-$L_3$ edge (approximately 2827-2847~eV). During these measurements, the sample and detector angles were adjusted simultaneously to maintain a fixed momentum transfer. For fixed-energy scans, on-resonance and off-resonance photon energies were selected at the energies corresponding to the maximum and minimum scattering intensity, respectively. The sample and detector angles were then varied to maintain a constant in-plane (lateral) scattering vector while systematically changing the out-of-plane component of the scattering vector. 

Figure \ref{Fig2}~(a) and (b) present the specular ($q_x$ = $q_y$ = 0) scattering intensity at a fixed $q_z$ =1.602 \AA $^{-1}$ measured at the strontium- and ruthenium-$L_3$ edges, respectively, near the 001$_{pc}$ Bragg peak of DyScO$_3$. The red and blue dots indicate the resonant and off-resonant energies, respectively, which were subsequently used for the measurements discussed below. By sitting at the on-resonance and off-resonance photon energies of each edge, we measured $q_z$ scans along the crystal truncation rod at different temperatures.  Figures~\ref{Fig2}(c) and \ref{Fig2}(d) show the out-of-plane diffraction profiles ($q_z$ scans) measured at the strontium-$L_3$ edge under resonant (1940~eV) and off-resonant (1930~eV) conditions at 170~K and 100~K, respectively. When the incident photon energy is tuned to the resonance, a pronounced enhancement and modulation of the oscillatory diffraction pattern is observed compared to the off-resonant case. The same is true for the ruthenium-$L_3$ edge. Figure~\ref{Fig2}(e) presents representative soft X-ray diffraction profiles measured at 170~K under resonant (2837~eV) and off-resonant (2827~eV) conditions. In both cases the, resonant scattering enhancements are on the higher $q_z$ and can be understood since the lattice parameter of SrRuO$_3$ at the DyScO$_3$ strain state leads to a Bragg peak at higher $q_z$ than DyScO$_3$. The estimated single-layer out-of-plane lattice constants are $c \approx 4.02~$\AA  ~for PbTiO$_3$ and $c \approx 3.89~~$\AA~ for SrRuO$_3$, as marked (Fig. ~\ref{Fig2}(c)). For both edges, there is no observable change in the region of 100-170 K (Supplementary Fig. S4 and S5), which crosses the ferromagnetic to paramagnetic transition \cite{10.1063/5.0087791}.

\begin{figure}[!ht]
\centering
\includegraphics[width=0.55\textwidth, height=0.5\textheight, keepaspectratio]{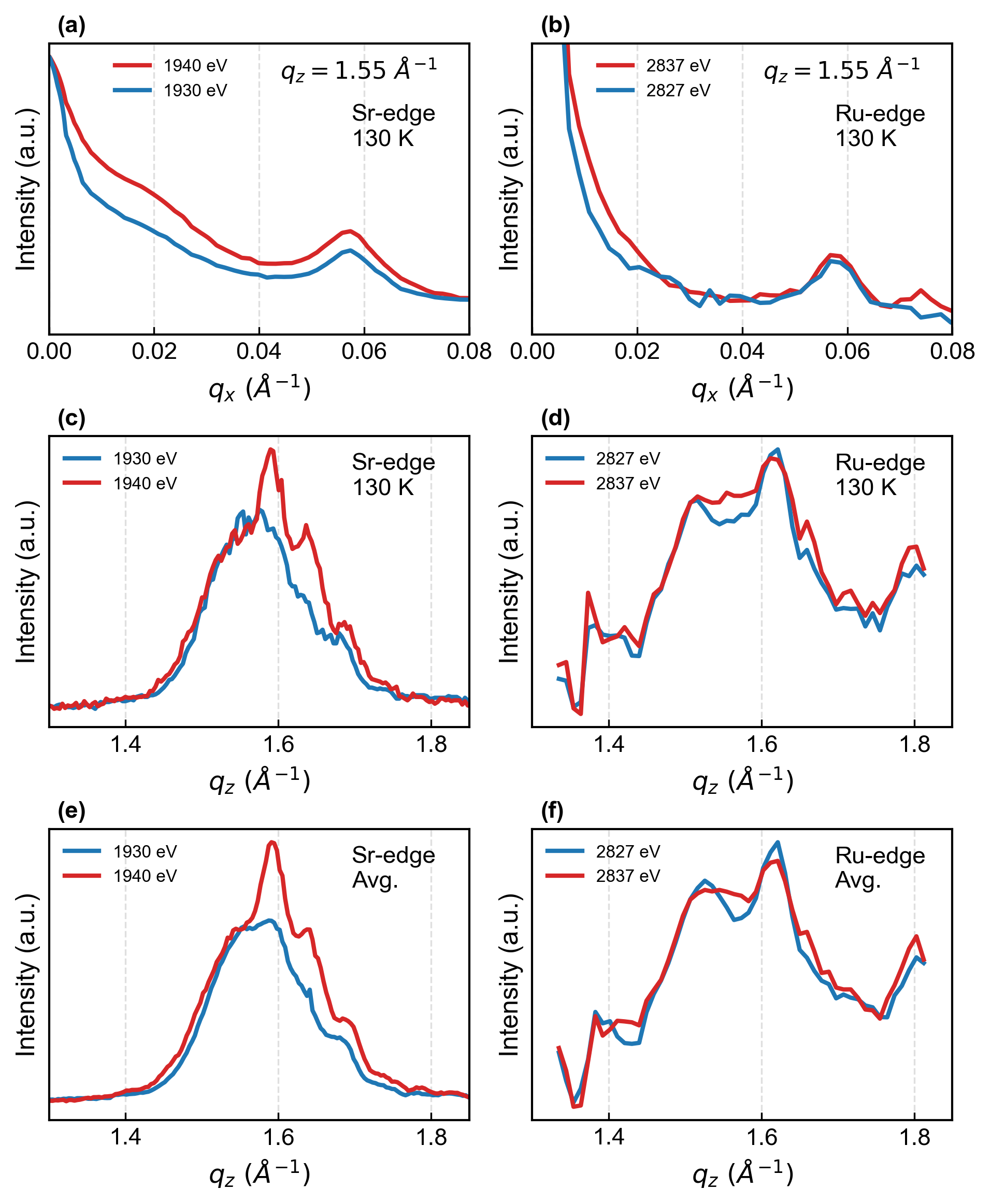}
\caption{\label{Fig3} In-plane $q_x$ scans of the X-ray reflectivity measured at a fixed $q_z$ corresponding to the PbTiO$_3$ layer, acquired at resonant and off-resonant energies at the strontium-$L_{3}$ edge at 130~K (a), In-plane $q_x$ scans of the X-ray reflectivity measured at a fixed $q_z$ corresponding to the PbTiO$_3$ layer, acquired at resonant and off-resonant energies at the ruthenium-$L_{3}$ edge at 130~K (b). Off-specular and out-of-plane soft X-ray reflectivity measured at 130~K at the strontium $L_{3}$ edge (c) and the ruthenium-$L_{3}$ edge (d) for the (PbTiO$_3$)$_{16}$/(SrRuO$_3$)$_9$/(PbTiO$_3$)$_{16}$ trilayer. Panels (e) and (f) show the averaged out-of-plane soft X-ray reflectivity over the temperature range 100–170~K at the strontium-$L_{3}$ and ruthenium-$L_{3}$ edges, respectively.}
\end{figure}

To understand the in plane lattice modulations, Figure~\ref{Fig3}(a) shows the resonant in-plane scattering intensity ($q_x$ scan) measured at the strontium-$L_3$ edge under resonant and off-resonant conditions at 130~K. Vortex satellite peaks are clearly observed when $q_z \approx 1.55~ $\AA$^{-1}$ is fixed at the PbTiO$_3$ lattice parameter is 4.02 \AA , consistent with the hard X-ray diffraction results, and confirming the presence of polar vortices. A pronounced change in the satellite peak intensities is observed upon tuning the photon energy from off-resonance to resonance. Similarly to strontium-\textit{L}$_3$ edge, the ruthenium-\textit{L}$_3$ edge resonant versus non-resonant measurement across the satellite peaks is enhanced for the former case (Fig.~\ref{Fig3}(b)). To see these enhancements more clearly, we fixed $q_x = 0.058$~\AA$^{-1}$, corresponding to the vortex satellite position, and performed $q_z$ scans crossing the vortex satellite peak to investigate the coupling to the structure of SrRuO$_3$ layer. 

The enhancement of the diffraction signal is observed when tuning the photon energy to the resonant conditions at both the strontium and ruthenium-$L_3$ edges when compared to the off-resonant case (Fig.~\ref{Fig3}(c, d)). However, the enhancement is less obvious at the ruthenium edge, compared to strontium. For off-specular scans, the detector also captures a background signal arising from X-ray fluorescence in addition to the scattered intensity; this contribution was carefully estimated and subtracted to isolate the true scattering signal.
Under resonant conditions, the measurement becomes element-specific, enhancing sensitivity to structural correlations associated with strontium and ruthenium. The diffraction profile exhibits a combination of sharp and broad features, where the sharp peaks indicate coherent, well-ordered regions across the layers, while the broader features reflect regions with reduced structural correlation length.

As in the specular case, these measurements were also performed over a broader temperature range from 100~K to 170~K in steps of 10~K (Supplementary Fig. S6). As no systematic temperature difference between scattering patterns was observed within this range, the datasets were averaged to improve statistical accuracy (Fig.~\ref{Fig3}(e) and \ref{Fig3}(f) for the strontium- and ruthenium-$L_3$ edges, respectively). At the strontium-$L_3$ edge (1940~eV), resonant soft X-ray scattering reveals a pronounced modulation at the out-of-plane Bragg peak position of the SrRuO$_3$ layer (Fig.~\ref{Fig3}(c)). Clear oscillations are observed with a periodicity of $\approx$ 0.048~\AA$^{-1}$, indicative of Laue oscillations with a corresponding coherence length slightly smaller than the entire thickness of the heterostructure.
At the ruthenium-\textit{L}$_3$ edge, the intensity of coherent oscillation is reduced, and the broad peak less sensitive to the resonance dominate. Moreover, the FWHM of the satellite peak at the strontium edge is $\approx$ 0.1~ \AA$^{-1}$, much smaller than the FWHM at the ruthenium edge of $\approx$~0.2 \AA$^{-1}$. These combined observations support the observations from scans along the crystal truncation rod (Fig.~\ref{Fig2}) and indicate that the coherent scattering on the strontium sub-lattice is much better defined compared to the one for the ruthenium sub-lattice, supporting relative disorder for the latter.

As we model below, this shows that the vortex strain modulations have contributions from strontium and ruthenium. This resonant enhancement demonstrates that the vortex-induced modulation penetrates into the SrRuO$_3$ layer, indicating that lattice distortions associated with the polar-vortex order in the PbTiO$_3$ layers extend across the interface. These modulations are attributed to strain originating from the vortex structure in the PbTiO$_3$ layers and are extended into the SrRuO$_3$ layer, which can only happen if the strain penetrates from PbTiO$_3$ into SrRuO$_3$.

\section{III. Diffraction Modeling}
\label{Diffraction}
\subsection*{A. Specular diffraction}

To better understand the data, diffraction simulations were performed based on a model of the heterostructure that takes into account layer dependent strains and with the first-order approximation of the structure as cubic (Supplementary note 1). First, we explore the case of the specular scattering. Note that given the mixed-polar phase with strong in-plane modulations, these are very challenging to simulate for the trilayer case. Such basic "epitaxial deformation diffraction model" (EDDM) simulations of the diffraction pattern near the 001-Bragg peak indicate that the origin of the peaks (marked in Fig. 2(c)) are due to the scattering of individual layers and interference effects within the heterostructure. The exact position of the peaks depends on the thickness of the layers, strain state, and structure factors of heterostructure and substrate. Due to the lattice mismatch between the PbTiO$_3$ and SrRuO$_3$ layers and the DyScO$_3$ substrate (Supplementary Note 1), we expect the scattering at larger q$_z$ than the Bragg peak to be more sensitive to SrRuO$_3$. Using EDDM, we simulated three different cases: (i) a single layer of PbTiO$_3$ (16 unit cells) on a DyScO$_3$ substrate (Supplementary Fig. S1), (ii) a bilayer (PbTiO$_3$)$_{16}$/(SrRuO$_3$)$_9$, and (iii) a trilayer system similar to the experimental heterostructure (PbTiO$_3$)$_{16}$/(SrRuO$_3$)$_9$/(PbTiO$_3$)$_{16}$ (Supplementary Fig. S2).

For the single-layer sample, a clear 16 u.c. periodicity is observed. In the bilayer case, two periodicities appear: one corresponding to the total thickness of 25 u.c., and another corresponding to the 9 u.c. SrRuO$_3$ layer. In contrast, the trilayer heterostructure exhibits a more complex periodicity that is difficult to quantify due to the enhanced structural modulation. In addition, the coexistence of mixed phases and the reported lattice-spacing variation \cite{stoica2019optical}. Substantially increases the number of free parameters, leading to strong parameter correlations and non-uniqueness in the fitting. Consequently, although extensive fitting trials were performed, a robust and physically meaningful fit to the specular peak measurements could not be obtained within reasonable constraints. 

\subsection*{B. Strain and polarization from phase-field simulations}

To understand the off-specular data, we need a model of the in-plane lattice from which we can calculate the satellite scattering. Following previous work on (PbTiO$_3$)$_n$/(SrTiO$_3$)$_n$ superlattices grown on DyScO$_3$(110) substrates, phase-field simulations were performed to model the polarization and strain distributions\cite{stoica2019optical, hong2017stability, damodaran2017phase, li2021subterahertz}. The resulting deformation pattern was projected on an atomic lattice to simulate X-ray diffraction \cite{yang2023thermodynamics, yang2022computing}. In order to understand the scattering, however, we need to project the strain patterns imposed by vortices in PbTiO$_3$ on the SrRuO$_3$ layer rather than SrTiO$_3$ as in prior reports. In order to compare modeling with experiment that utilize already established vortex structures in (PbTiO$_3$)$_n$/(SrTiO$_3$)$_n$ (see Supplementary note 2 and Fig. S3), we choose to utilize the (PbTiO$_3$)$_n$/(SrTiO$_3$)$_n$ data as a proxy for the strain in the SrRuO$_3$ layer and asses its relevance in resonant diffraction simulations. We note that in SrTiO$_3$, the polarization magnitude is much smaller than in PbTiO$_3$ and the primary effect is the lattice strain in SrTiO$_3$. The resulting strain distributions at the interface were then used as input for subsequent atomistic calculations, in which the SrTiO$_3$ layer was replaced by SrRuO$_3$. This approach enables a direct assessment of the influence of interfacial strain modulation on the electronic and magnetic properties of SrRuO$_3$ while preserving the experimentally relevant strain environment. While the quantitative value of the strain in SrRuO$_3$ will be different, the patterns should be similar.
\begin{figure}[!ht]
\centering
\includegraphics[width=0.88\textwidth]{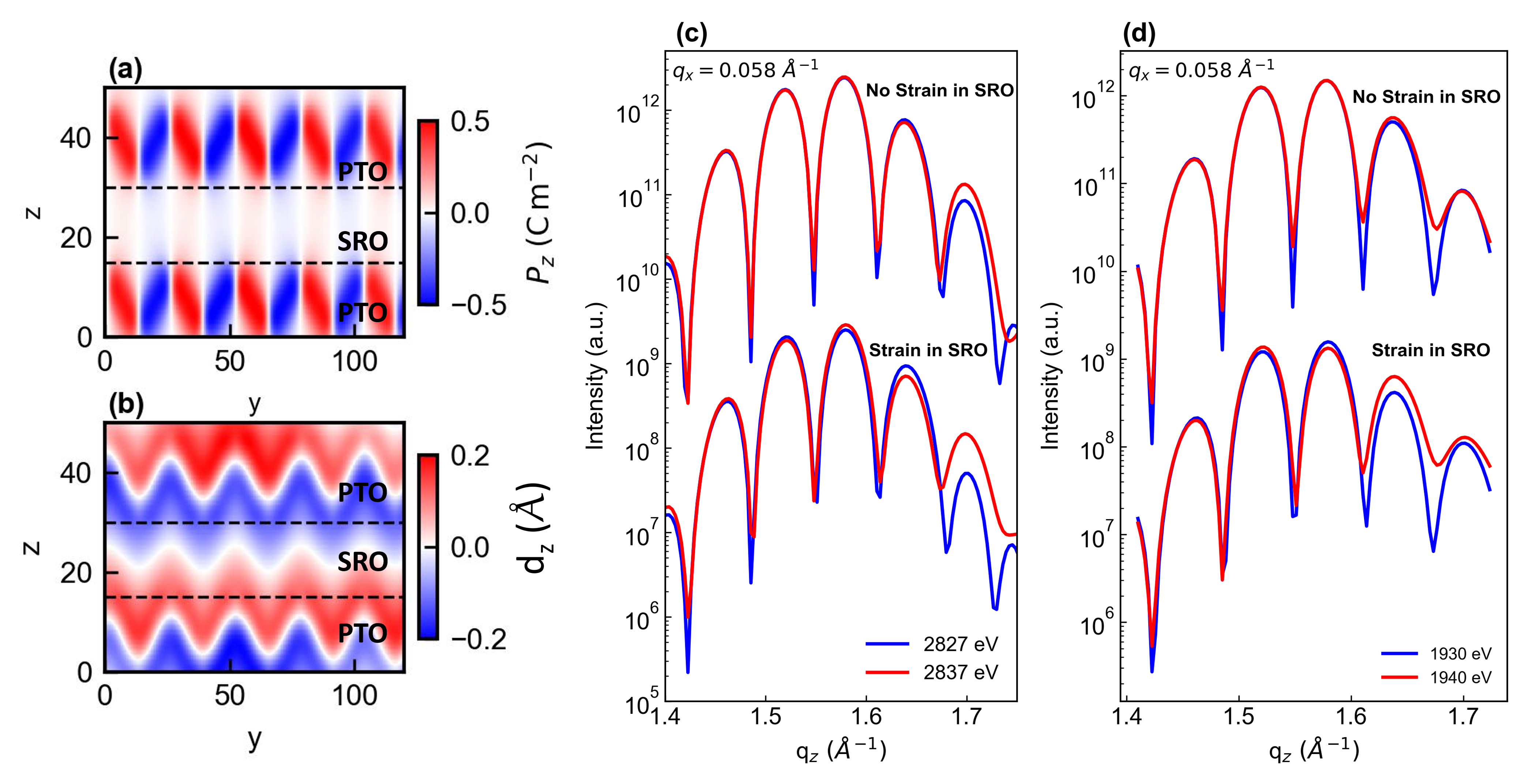}
\caption{\label{SimRSM} Effects of epitaxial strain on resonant X-ray scattering, phase-field simulations yield the out-of-plane polarization $P_z$ (a) and mechanical displacement $d_{z}$ (b). Simulated off-specular out-of-plane $q_z$ reflections measured at resonant and off-resonant energies for ruthenium edge (c), and strontium edge (d), comparing the cases where strain is absent in the SrRuO$_3$ layer and where strain is active in the SrRuO$_3$ layer, are shown in top and bottom panels, respectively. 
}
\end{figure}
\
\begin{figure}[!ht]
\centering
\includegraphics[width=0.8\textwidth]{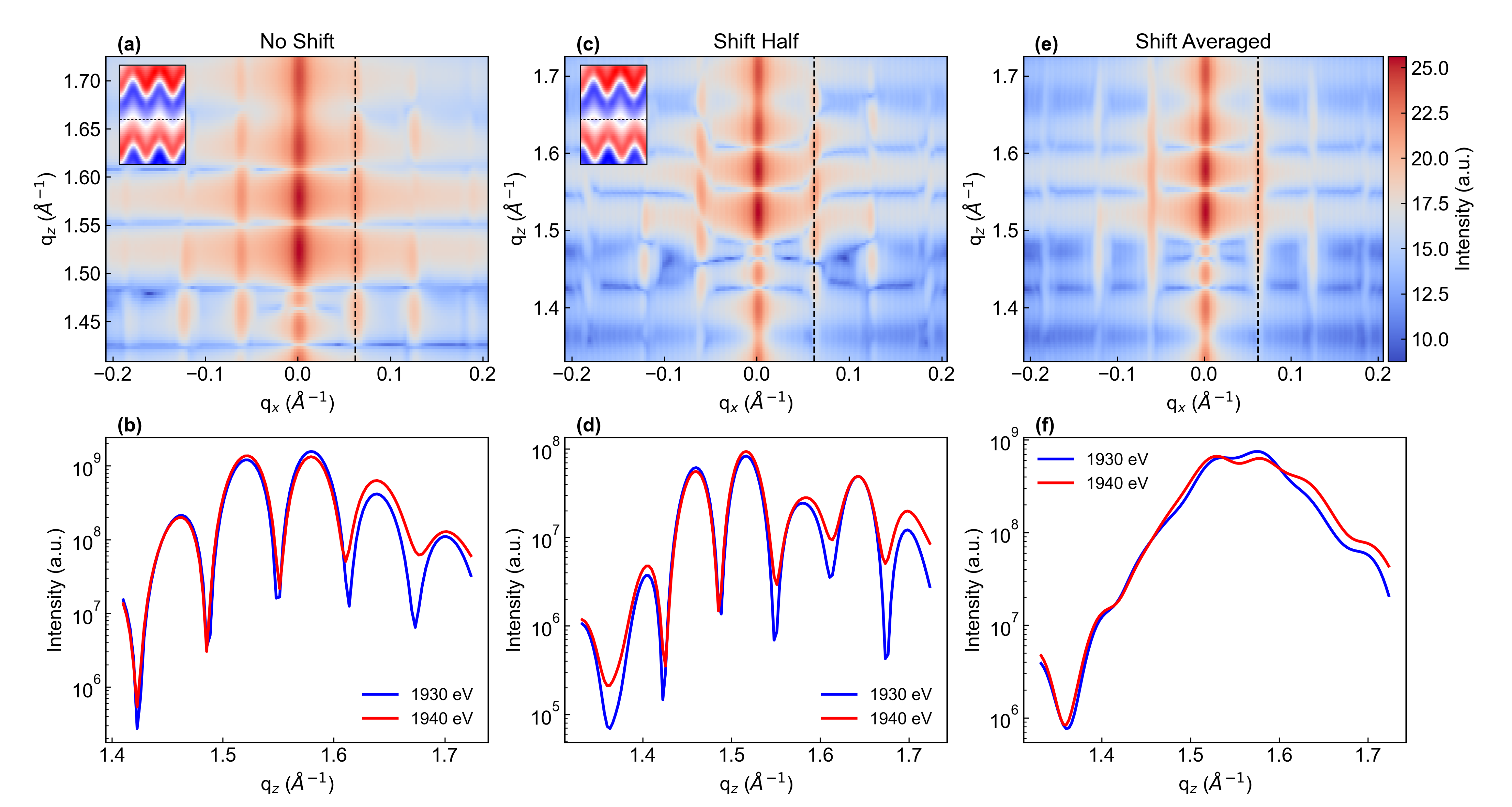}
\caption{\label{SimRSM1} Effects of interfacial vortex displacement on resonant X-ray scattering, simulated reciprocal space map (RSM) of the trilayer heterostructure, along with the corresponding vertical cuts through the off-specular peaks, is shown for different cases: no atomic displacement shift (a, b), half-period atomic displacement shift (c, d), and the average of all small-step displaced atomic positions (e, f). Off-specular out-of-plane 
\textit{q}$_z$ reflections were measured at both resonant and off-resonant energies. Black dotted lines indicate the vertical cuts corresponding to the off-specular diffraction peaks.
}
\end{figure}

Figures~\ref{SimRSM}(a) and \ref{SimRSM}(b) show the out-of-plane polarization and mechanical displacement components along the $z$ axis, $P_z$, obtained from phase-field simulations in the vortex region of the sample. First, this vortex region leads to vertical strain modulations that are observable along the crystal truncation rod near the 001-Bragg peak. However, the in-plane modulations in the \textit{a}$_1$/\textit{a}$_2$ regions of the sample (Supplementary Fig. S7) also contribute to the scattering along the crystal truncation rod (H = K = 0). If we move in the off-specular region of the diffraction pattern, a peak with H $\neq$ 0 or K $\neq$ 0 for \textit{a}$_1$/\textit{a}$_2$ are not accessible at the strontium edge around the 001 Bragg peak (since \textit{a}$_1$/\textit{a}$_2$ satellites for are not observed around the specular peak) and difficult to measure at the ruthenium edge around other Bragg peaks due to the low flux of the beamline in this energy region. In the vortex region of the sample, the satellite peaks due to both in-plane polarization and strain modulations are present around the 001 Bragg peak. The vortex-driven strain is primarily out of plane (cite supplement) and key for the interpretation of experimental diffraction patterns as discussed below. 

To simulate off-specular diffraction, we use (PbTiO$_3$)$_{16}$/(SrRuO$_3$)$_9$/(PbTiO$_3$)$_{16}$ in the $z$ direction and repeat this to make a lattice structure N unit cells wide (along y-axis), where N = 500. For this case, we remove the substrate contributions to simplify the calculation. From this structure, off-specular $q_z$ diffraction profiles calculated at the ruthenium edge (Fig.~\ref{SimRSM}(c)) and strontium edge (Fig.~\ref{SimRSM}(d)) are shown for two limiting cases: (i) absence of strain in the SrRuO$_3$ layer (upper panel) and (ii) presence of strain in the SrRuO$_3$ layer (lower panel). The pronounced contrast between these two scenarios demonstrates that the experimentally observed resonant features are consistent with strain-mediated coupling within the SrRuO$_3$ layer. Moreover, the simulations assume an idealized, defect-free structure; consequently, the calculated RSMs exhibit sharper features, including a pronounced dip, than those observed experimentally due to film thickness variations. It is worth noting that the calculated resonant enhancements are less pronounced than those observed experimentally, while the inclusion of strain-driven lattice displacements considered in simulations takes the leading role in capturing the resonant feature of the experiment. The smaller resonant changes observed in simulations relative to experiment may arises due to the absence of DyScO$_3$ substrate in diffraction simulations, which was not considered for simplicity, but may play a role to enhance resonant scattering in experiment via multiple reflections and interference effects. The DyScO$_3$ substrate (Supplementary Fig.~S1) itself introduces an asymmetry between lower and higher $q_z$ values, which contributes to the enhanced peak asymmetry observed in the experimental data.

In the simulations (Fig.~\ref{SimRSM}), we assumed perfect alignment of the vortices between the two layers.  TEM experiments, however, show that there is often disorder in alignment between features in adjacent PbTiO$_3$ layers \cite{abid2021creating}. To examine the impact of this disorder, we separated the trilayer system into two halves consisting of (PbTiO$_3$)$_{16}$/(SrRuO$_3$)$_9$/(PbTiO$_3$)$_{16}$. We then shifted the strain and polarization profiles of one half laterally at a series of values over the repeat distance of the vortex structure. The simulated RSM at 1930~eV and the corresponding $q_z$ cuts under resonant and off-resonant conditions for the case without atomic displacement are shown (Fig.~\ref{SimRSM1}(a, b)). A clear enhancement is observed under resonant conditions on the higher $q_z$ side. Next, we introduced an out-of-phase atomic displacement in the upper half of the trilayer (Fig.~\ref{SimRSM1}(c, d)). In this case, tilted satellite peaks emerge as a consequence of the out-of-phase modulation, and the resonant enhancement along $q_z$ becomes less pronounced compared to the no-displacement case.
Finally, we displaced the atomic positions in finite steps and averaged over all displaced configurations. The resulting RSM and corresponding $q_z$ cuts are shown (Fig.~\ref{SimRSM1}(e, f)). In this averaged case, a broader diffraction feature is observed, with no significant enhancement on the lower $q_z$ side and a remaining enhancement on the higher $q_z$ side.

\section{IV. Discussion}
\label{sec:discussion}
Having established that reciprocal space patterns near the strontium and ruthenium resonance are connected with the real space model structure of the (PbTiO$_3$)$_{16}$/(SrRuO$_3$)$_9$/(PbTiO$_3$)$_{16}$ trilayer allows us to discuss the overall results in greater depth. Our discussion focuses on analyzing how the polar vortex influences strain modulation inside the (PbTiO$_3$)$_{16}$/(SrRuO$_3$)$_9$/(PbTiO$_3$)$_{16}$ trilayer, specifically examining how this effect propagates from the ferroelectric PbTiO$_3$ regions into the ferromagnetic SrRuO$_3$ layer.   

We have consistently observed that for $q_z >$ 1.6 \AA  $^{-1}$, the experimental resonant scattering at the strontium and ruthenium characteristic energies, compared to the non-resonant reference scans, is enhanced on both specular (Fig.~\ref{Fig2}) and off-specular diffraction (Fig.~\ref{Fig3}). This contrasts with the specular and off-specular scattering pattern at $q_z <$ 1.6 \AA  $^{-1}$, where the resonant scattering enhancement is much weaker or even absent. These results can be tracked to the strain modulation forming inside the SrRuO$_3$ layer coupled to the periodicity of the vortex structure in the adjacent PbTiO$_3$ layers. To quantitatively support the strain origin of the resonant scattering enhancement at $q_z >$ 1.6 \AA  $^{-1}$, the diffraction modeling based on the phase-field simulation of the real-space structure quantified the strain modulation simultaneously present in the PbTiO$_3$ and SrRuO$_3$ layers to decide that resonant-scattering enhancement is dominated by strain modulation (Fig.~\ref{SimRSM}, and Fig. S8). While full quantitative agreement with experimental data is still challenging, the favorable comparison of the main experimental trends relative to the simulations provide confidence to the assignment of such effects to shared strain modulations between the layers.

As we captured both coherent and incoherent stacking features in the experimental diffraction patterns, the comparison between measurements at the ruthenium and strontium $L_3$ edges reveals differences in the character of the oscillatory features, reflecting variations in lattice coherence across the heterostructure. The diffraction profiles at the strontium and ruthenium edges shows\ a broader distribution of features alongside sharper components, suggesting a coexistence of regions with varying degrees of structural correlation. The strontium edge, however, exhibits more pronounced oscillatory behavior, consistent with a more extended penetration of strontium sub-lattice coherence compared to the ruthenium one. These observations are consistent with the trends presented in Fig.~\ref{SimRSM1}, where phase-field simulations captured the broad peak observed in experiments (Fig.~\ref{Fig3} c, f) by accounting for lattice incoherency between the top and bottom parts of the heterostructure (Fig.~\ref{SimRSM1} e, f). This is contrast to the sharp resonant enhancement of coherent oscillations observed at the strontium resonance (Fig.~\ref{Fig3} a, c, e) that is captured in the coherent lattice simulation of heterostructure (Fig.~\ref{SimRSM1} a, b). These comparisons allows us to identify that in the experimental sample coherent and incoherent regions coexist in the heterostructure competing in the phase ordering space. Further refinements of the real-space structure can include the DyScO$_3$ substrate and microstructure variations \cite{shao2023real} as well as more detailed experimental tuning (resonant conditions and multi-peak determinations) to match a multitude of simulated and experimental variables is left for future work.

Next, we compare our results with related studies in literature. High-resolution transmission electron microscopy directly determined the atomistic structure of similar vortex textures with $\approx$ 8.5 nm periodicity in \cite{rusu2022ferroelectric} in SrRuO$_3$/PbTiO$_3$/SrRuO$_3$ trilayers, the direct observation of SrRuO$_3$ lattice modulation connected with a magnetic probe remains challenging at this length scale \cite{seddon2021real}. Furthermore, SrRuO$_3$/PbTiO$_3$/SrRuO$_3$ trilayers with variable PbTiO$_3$ thickness indicated a critical thickness of $>$~ 23 u.c. needed to penetrate strain modulation throughout the entire heterostructure \cite{lichtensteiger2023nanoscale}, thus ruling out the coherency for in-plane strain modulations across heterostructure interfaces linked with vortex structures that are stable at $<$~25 u.c. \cite{hong2021vortex}. In another related study of PbTiO$_3$/Pb$_{1-x}$Sr$_x$TiO$_3$ heterostructures with a thickness comparable to our sample, the PbTiO$_3$ formed vortices, while the Pb$_{1-x}$Sr$_x$TiO$_3$ formed its own $a_1/a_2$  twin structures with different periodicity and nanodomain wall anisotropy \cite{kavle2024highly}, thus indicating decoupling of layers in contact with vortex structures in PbTiO$_3$.  

Thus, our experiments address a challenge in probing penetrating strain modulations with $\approx$ 10 nm across interfaces of PbTiO$_3$/SrRuO$_3$ heterostructures using resonant X-ray scattering techniques with element specificity. Moreover, tuning the stacking order of vortex structures in the top and bottom PbTiO$_3$ layers shows that the incoherent structure is suppressing the coherent resonant oscillations at the resonance (Fig.~\ref{SimRSM1} (e,f)), while the experiment on the contrary (Fig.~\ref{Fig3} (d, f)) enhances the coherent oscillations observed at the resonance, to attest that the experimental structure indeed possesses penetrating collective strain modulations though the SrRuO$_3$, with possible implications for future studies on controlling magnetic heterogeneity at a length scale approaching a few unit cells. Such magnetic heterogeneity could be probed through anomalous Hall effect measurements, as suggested in recent studies \cite{seddon2021real, nishihaya2025spontaneous}. In addition, X-ray magnetic circular dichroism (XMCD) could also provide complementary, element-specific insight into these magnetic modulations through their local magnetic response. However, XMCD measurements were not performed in the present work because the magnetic field required to fully saturate the magnetic moment in SrRuO$_3$ exceeds the maximum field available at our XMCD endstation.

While it is well established that lattice strain is a powerful tuning parameter for magnetic properties in complex oxides, capable of modifying exchange interactions, magnetic anisotropy, and spin-orbit coupling~\cite {koster2012structure}, enabling control of magnetic properties on the typical length scales of magnetic domains, typically in the range of $>$ 100 nm. The spatially varying strain fields with $\approx$ 10 nm periodicity provided by polar vortices at SrRuO$_3$ interfaces are particularly significant because the associated strain gradients can give rise to local variations in magnetic interactions, including anisotropy, exchange interaction, and Dzyaloshinskii-Moriya interactions. These effects are known to stabilize chiral magnetic textures \cite{wang2020magnetic,huang2020detection}, but with smaller dimensions based on heterostructure strain interactions. By transferring vortex-driven strain into the SrRuO$_3$ layer, the nanoscale periodic modulation demonstrated here provides a natural platform for engineering spatially ordered magnetic textures that are inaccessible using conventional strain engineering, highlighting their potential for next-generation spintronic applications.

\section{Conclusion}
\label{sec:concl}
In conclusion, we combined nonresonant hard X-ray and resonant soft X-ray scattering to investigate polar-vortex–driven interfacial strain coupling in (PbTiO$_3$)$_{16}$/(SrRuO$_3$)$_9$/(PbTiO$_3$)$_{16}$ heterostructures. While non-resonant hard X-ray diffraction reveals a phase mixture of $a_1/a_2$ ferroelectric domains and polar vortex structures in the heterostructures, we employed soft X-ray probes of vortex structures using diffuse-scattering peaks probed at the \textit{L}$_3$ edges of strontium and ruthenium to reveal vortex-strain modulations superimposed on the SrRuO$_3$ layers. Electrical transport measurements confirm that the SrRuO$_3$ layer remains ferromagnetic below its Curie temperature, establishing a robust magnetic ground state in the heterostructure geometry. Element-specific resonant soft X-ray scattering at the strontium- and ruthenium-$L_3$ edges reveals pronounced intensity modulations, demonstrating that the polar vortex supertextures in PbTiO$_3$ are transferred into the adjacent SrRuO$_3$ layer. A resonant-diffraction model based on phase-field simulations of the real-space structure accurately reproduces the key features of the experimental reciprocal space maps and attributes the strong resonant enhancement near the off-specular Bragg condition to strain-mediated lattice modulations that penetrate the SrRuO$_3$. Taken together, these results establish strain transfer as an effective mechanism for imprinting polar supertextures onto a ferromagnetic layer, offering a pathway for interrogating nanoscale magnetic textures coupled to lattice modulations in oxide heterostructures at length scales $<$ 10 nm. The observation of nanoscale modulated strain interaction between ferromagnetic layers placed in contact with ordered polar vortex arrays open the perspective to investigate vortex state in multiferroic regimes, extending its functional properties.

\section{Acknowledgements}
 \begin{acknowledgements}   This work was primarily supported by the U.S. Department of Energy, Office of Science, Office of Basic Energy Sciences, under Award Number DE-SC-0012375 for work with ultrafast X-Ray and optical experiments. L.W.M. additionally acknowledges the support of the Army Research Office under the ETHOS MURI via cooperative agreement W911NF-21-2-0162 for the development of the materials and heterostructures. Work at the Advanced Photon Source, Argonne was supported by the U.S. Department of Energy, Office of Science under Grant No. DEAC02-06CH11357.  Work performed at the Center for Nanoscale Materials, a U.S. Department of Energy Office of Science User Facility, was supported by the U.S. DOE, Office of Basic Energy Sciences, under Contract No. DE-AC02-06CH11357.
 
 \end{acknowledgements}

\bibliography{Trilayer.bib}

\end{document}